\documentclass{article}
\usepackage{spconf,amsmath,graphicx}

\usepackage{amsmath}
\usepackage{nccmath}
\usepackage{cite}
\usepackage{amssymb}
\newcommand{\blskip}{\baselineskip   10.15pt}
\newcommand{\ftblskip}{\baselineskip 10.15pt}

\def\X{{\mathbf{X}}}

\def\0{{\mathbf{0}}}

\def\sos{\mathtt{<\!\!sos\!\!>}}
\def\eos{\mathtt{<\!\!eos\!\!>}}
\def\Pe2e{P_{\mathtt{e2e}}}

\def\Pdec{P_{\mathtt{dec}}}
\def\Pflm{P_{\mathtt{flm}}}
\def\Pblm{P_{\mathtt{blm}}}

\title{ITERATIVE SHALLOW FUSION OF BACKWARD LANGUAGE MODEL\\FOR END-TO-END SPEECH RECOGNITION}
\name{Atsunori Ogawa, Takafumi Moriya, Naoyuki Kamo, Naohiro Tawara, and Marc Delcroix}
\address{NTT Corporation, Japan}
\begin{document}
\ninept
\blskip
\maketitle
\begin{abstract}
We propose a
new
shallow fusion (SF) method
to
exploit
an external backward
language model (BLM)
for end-to-end automatic speech recognition (ASR).
%
The
BLM has
complementary characteristics
with a forward
language model (FLM),
and the effectiveness of their combination has been confirmed
by
rescoring ASR hypotheses
as post-processing.
In the proposed SF,
we iteratively apply the BLM
to
partial
ASR
hypotheses
in the backward direction
(i.e.,
from the possible next
token
to the
start symbol)
during decoding,
substituting the newly calculated BLM scores
for the scores calculated at the last iteration.
To enhance the effectiveness
of
this
iterative SF (ISF),
we train a
partial sentence-aware BLM (PBLM)
using
reversed
text data
including
partial sentences,
considering
the framework
of ISF.
In experiments using
an attention-based encoder-decoder ASR system,
we confirmed that
ISF using the PBLM shows
comparable performance
with SF using the FLM.
By performing ISF,
early
pruning of prospective
hypotheses can be
prevented
during decoding,
and
we
can obtain a performance improvement
compared
to
applying
the PBLM
as
post-processing.
Finally,
we
confirmed that,
by combining SF and ISF,
further performance improvement can be obtained
thanks to the complementarity of the FLM and PBLM.
\end{abstract}
%
\section{Introduction}
\label{sec_intro}
%
Thanks to the
significant
progress
in
neural network (NN) modeling,
the performance of automatic speech recognition (ASR)
has
rapidly improved.
%
It was reported in 2017 that
a hybrid ASR system,
which exploits
neural
acoustic modeling
and neural language model (LM) rescoring
(and
also
system combination),
had surpassed
human performance
on the conversational speech recognition task
\cite{Xiong_arXiv2017,Xiong_IEEEACMTASLP2017,Saon_IS2017}.
%
The performance
has been further improved by
leveraging a fully neural end-to-end (E2E) architecture
\cite{Tuske_IS2020,Tuske_IS2021}.

The E2E ASR system
can integrate
an external
LM
trained
using 
a large text corpus
to
improve the system performance
\cite{Tuske_IS2020,Tuske_IS2021}.
Various
methods for
integrating
the external
($n$-gram or neural)
LM
have been proposed
\cite{Gulcehre_arXiv2015,Bahdanau_ICASSP2016,Chorowski_IS2017,Hori_IS2017,Kannan_ICASSP2018,Toshniwal_SLT2018,Sriram_IS2018,Shan_ICASSP2019,McDermott_ASRU2019}.
Shallow fusion (SF)
\cite{Gulcehre_arXiv2015,Bahdanau_ICASSP2016,Chorowski_IS2017,Hori_IS2017,Kannan_ICASSP2018,Toshniwal_SLT2018}
is the most popular method that
performs
log-linear interpolation between
scores
obtained from the main
ASR model and
the
external LM during decoding.
SF
is
simple,
but
it
shows comparable performance
with the learning-based
deep
\cite{Gulcehre_arXiv2015},
cold
\cite{Sriram_IS2018},
and component
\cite{Shan_ICASSP2019}
fusion methods,
which jointly
fine-tune
the
main
ASR model and the pre-trained external neural LM
for their tighter integration
\cite{Toshniwal_SLT2018,Shan_ICASSP2019,McDermott_ASRU2019}.
The more recently
proposed
density ratio approach
\cite{McDermott_ASRU2019}
and internal language model estimation
\cite{Meng_SLT2021,Moriya_ICASSP2022}
intend to reduce the influence of the internal LM
implicitly included in the
main
ASR model
to enhance the effect of
LM integration.
%
These methods
inherit
SF's
simplicity and practicality
(since
they can be
performed
during decoding without requiring any model training)
and outperform SF
\cite{Tuske_IS2021,McDermott_ASRU2019,Meng_SLT2021,Moriya_ICASSP2022}.
We focus on
simple
SF
in this study
since it shows good enough performance
\cite{Toshniwal_SLT2018}.

Since
ASR hypotheses are extended
successively
from the start of an input utterance
during decoding (beam search),
a forward
LM (FLM) is a natural choice for SF
\cite{Tuske_IS2020,Gulcehre_arXiv2015,Bahdanau_ICASSP2016,Chorowski_IS2017,Hori_IS2017,Kannan_ICASSP2018,Toshniwal_SLT2018}.
In contrast,
in post-processing,
such as $N$-best rescoring and lattice rescoring,
a backward
LM (BLM)
can
also
be
used along with the FLM
\cite{Xiong_arXiv2017,Xiong_IEEEACMTASLP2017,Irie_IS2018,Kanda_CHiME5,Arora_CHiME6,Ogawa_ICASSP2022}.
The BLM has complementary characteristics with the FLM,
and
their combination
can
greatly improve the ASR performance.
Besides,
the
post-processing
is performed for completed
(not partial)
hypotheses
after decoding,
and
the BLM can be easily applied to the hypotheses
from their end-of-sequence ($\eos$) symbols.
In other words,
it is difficult to apply the BLM for partial hypotheses during decoding,
since they do not have the $\eos$ symbols,
and this can limit the effect of the BLM.

In this study,
we propose a new SF method to exploit an external BLM.
To the best of our knowledge,
this is the first study to use the BLM in SF.
We
expect that,
by applying the BLM
along with an FLM 
during decoding
(not as post-processing),
more accurate LM scores can be given
to partial ASR hypotheses,
and thus we can prevent incorrect early
beam
pruning of
the prospective hypotheses.
Consequently,
we can obtain a more accurate one-best hypothesis
as the ASR result.
In contrast to
the
conventional SF
using the FLM,
which
can
be performed
cumulatively,
SF using the BLM needs to be performed iteratively.
When
a possible next token is connected
to a partial ASR hypothesis
during decoding,
we need to recalculate
the BLM score
for
the
entire
hypothesis
in the backward direction
(i.e., from the possible next token to the start symbol)
and substitute it for the score calculated at the last iteration.
Moreover,
to
enhance the effectiveness
of the proposed
iterative SF (ISF),
we train a
partial sentence-aware BLM
using
reversed
text data
that includes
partial sentences
considering
the framework
of ISF.
%
%
We conduct 
experiments
on the TED talk corpus
\cite{Rousseau_LREC2014}
using
an attention-based encoder-decoder
(AED)
ASR system
\cite{Watanabe_IEEEACMTASLP2017,Guo_ICASSP2021,ESPnet}
as the E2E ASR system
and
long short-term memory (LSTM)-based
recurrent NN
LMs (LSTMLMs)
\cite{Hochreiter_NeuralComput1997,Sundermeyer_IS2012}
as the external LMs.
We
show
the effectiveness of the proposed ISF
and
the effectiveness
of its combination with the conventional SF.
%
\vspace{-2.25mm}
\section{Iterative shallow fusion}
\label{sec_prop}
\vspace{-1.50mm}
%
We describe the framework of the proposed ISF,
the computational cost reduction of ISF-based decoding,
and
a partial sentence-aware BLM
suitable for ISF.
%
\vspace{-3.00mm}
\subsection{Framework of ISF}
\label{ssec_fw}
\begin{figure}[t!]
  \centering
  \includegraphics[width=0.918\linewidth]{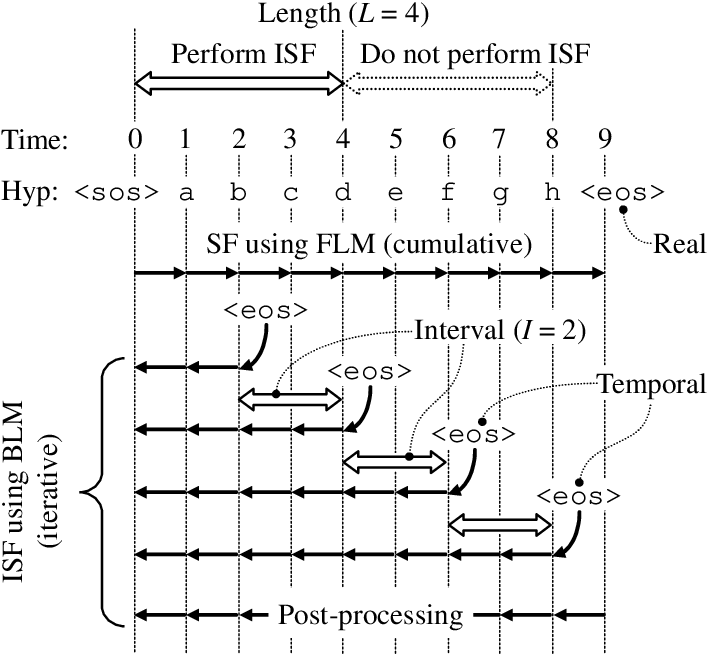}
  \vspace{-2.25mm}
  \caption{\ftblskip Framework of the proposed ISF using a BLM. $\sos$ and $\eos$ are the start and (temporal or real) end symbols of an ASR hypothesis (token sequence), respectively. An alphabet corresponds to a token (e.g., a word, a subword, or a character).}
  \label{fig_isf}
  \vspace{-4.75mm}
\end{figure}

\vspace{-1.00mm}
%
Figure~\ref{fig_isf} shows the framework of the proposed ISF using a BLM.
Let $\X$ be a hidden state vector sequence
for
an input utterance
obtained from the encoder of an
AED ASR model.
Given $\X$,
we perform decoding
with
SF using an FLM
and ISF
with
a BLM
on the decoder of the
AED ASR model.
Let
$\left\{\sos,w_{1:t-1}\right\}$
be a partial hypothesis (token sequence) of length $t-1$
and $w_{t}$ be a possible next token
connected to
$\left\{\sos,w_{1:t-1}\right\}$,
where $\sos$ is the start-of-sequence symbol
that
is excluded from the hypothesis length count.
We
calculate the score
(log probability)
of the extended partial hypothesis
$\left\{\sos,w_{1:t}\right\}$
given $\X$
as,
\begin{subequations}
\begin{fleqn}[5.00mm]
\label{eq_isf}
\begin{align}
\lefteqn{\log{P(\left\{\sos,w_{1:t}\right\}|\X)}}
\nonumber \\
&
=\log{P(\left\{\sos,w_{1:t-1}\right\}|\X)}
\label{eq_isf_rterm1} \\
&\hspace{3.60mm}
+\log{\Pdec(w_{t}|\X,\left\{\sos,w_{1:t-1}\right\})}
\label{eq_isf_rterm2} \\
&\hspace{3.60mm}
+\alpha\log{
\Pflm(w_{t}|\left\{\sos,w_{1:t-1}\right\})
}
\label{eq_isf_rterm3} \\
&\hspace{3.60mm}
+\beta
\left[
\hspace{0.25mm}
\log{\Pblm(\left\{\eos,w_{t:1},\sos\right\})}
\right.
\nonumber \\
&\hspace{6.20mm}
\left.
-
\hspace{1.00mm}
\log{\Pblm(\left\{\eos,w_{t-1:1},\sos\right\})}
\hspace{0.25mm}
\right]
\label{eq_isf_rterm4} \\
&\hspace{3.60mm}
+\,
\gamma,
\label{eq_isf_rterm5}
\end{align}
\end{fleqn}
\end{subequations}
where the first term on the right side of this equation
(Term~(\ref{eq_isf_rterm1}))
is the score
of
$\left\{\sos,w_{1:t-1}\right\}$
given $\X$,
Term~(\ref{eq_isf_rterm2})
is the score
of $w_{t}$ given $\X$ and
$\left\{\sos,w_{1:t-1}\right\}$
obtained from
the decoder
of
the E2E ASR model,
Term~(\ref{eq_isf_rterm3})
is the score
of $w_{t}$ given $\{\sos,w_{1:t-1}\}$
obtained by
the conventional
SF using the FLM
scaled by
weight $\alpha$ ($\alpha \geq 0$),
$\sos$ is used to start FLM calculation
in the forward direction,
Term~(\ref{eq_isf_rterm4})
is the score
obtained by
the proposed
ISF using the BLM
scaled by
weight $\beta$ ($\beta \geq 0$),
and the last
Term~(\ref{eq_isf_rterm5})
is
the
reward $\gamma$ ($\gamma \geq 0$)
given proportional to the sequence length
\cite{Tuske_IS2021,Bahdanau_ICASSP2016,McDermott_ASRU2019}.
The two BLM scores in
Term~(\ref{eq_isf_rterm4})
can be rewritten as,
\begin{fleqn}[5.00mm]
\begin{eqnarray}
\lefteqn{
\log{\Pblm(\left\{\eos,w_{t:1},\sos\right\})}}
\nonumber \\
&=&
\log{\Pblm(\eos)}
\nonumber \\
& &
+
\log{{\textstyle\prod_{\tau=t}^{1}}
\Pblm(w_{\tau}|\left\{\eos,w_{t:\tau+1}\right\})}
\nonumber \\
& &
+
\log{\Pblm(\sos|\left\{\eos,w_{t:1}\right\})},
\label{eq_isf_curr}
\end{eqnarray}
\end{fleqn}

\begin{fleqn}[5.00mm]
\begin{eqnarray}
\lefteqn{\log{\Pblm(\left\{\eos,w_{t-1:1},\sos\right\})}}
\nonumber \\
&=&
\log{\Pblm(\eos)}
\nonumber \\
& &
+
\log{{\textstyle\prod_{\tau=t-1}^{1}}
\Pblm(w_{\tau}|\left\{\eos,w_{t-1:\tau+1}\right\})}
\nonumber \\
& &
+
\log{\Pblm(\sos|\left\{\eos,w_{t-1:1}\right\})},
\label{eq_isf_last}
\end{eqnarray}
\end{fleqn}
where $\eos$ is the
temporarily connected
end-of-sequence symbol
used
to start BLM calculation
in the backward direction.

As can be seen in
Fig.~\ref{fig_isf} and Term~(\ref{eq_isf_rterm3}),
we can perform the conventional SF using the FLM
cumulatively.
At every timestep $t$,
we calculate the FLM score of $w_{t}$
conditioned on $\{\sos,w_{1:t-1}\}$,
and the calculated score of $w_{t}$ does not change
during decoding for the input utterance.
In contrast,
we cannot perform SF using the BLM
cumulatively.
The BLM score of $w_{\tau}$ changes
during decoding
when we extend the hypothesis with a new token,
i.e.,
the condition $\{\eos,w_{t:\tau+1}\}$
changes
at every timestep $t$,
as shown in Eq.~(\ref{eq_isf_curr}).

Therefore,
as shown in
Fig.~\ref{fig_isf} and Term~(\ref{eq_isf_rterm4}),
we need to perform SF using the BLM iteratively.
In the proposed ISF,
at every timestep $t$,
we
calculate the BLM score for
the
entire
partial hypothesis
$\{\eos,w_{t:1},\sos\}$,
as shown in Eq.~(\ref{eq_isf_curr}).
Furthermore,
since the BLM score at the last timestep (iteration) $t-1$
shown in Eq.~(\ref{eq_isf_last}) is included in
Term~(\ref{eq_isf_rterm1}),
we need to subtract the score from the term,
as shown in
Term~(\ref{eq_isf_rterm4}).

As shown in Fig.~\ref{fig_isf},
we also calculate the BLM score
for the hypothesis that reaches the real (not temporal) $\eos$,
i.e., for the completed (not a partial) hypothesis,
as a kind of post-processing.
However,
this post-processing is different
from
conventional post-processing
\cite{Xiong_arXiv2017,Xiong_IEEEACMTASLP2017,Irie_IS2018,Kanda_CHiME5,Arora_CHiME6,Ogawa_ICASSP2022}.
%
The conventional
post-processing is performed
for a limited number of
hypotheses
(e.g., $N$-best list) after decoding.
In contrast,
this post-processing is performed
for all the
hypotheses that reach $\eos$ during decoding.
%
Consequently,
this post-processing is richer
than
the
conventional post-processing.
%
%
\subsection{Reducing computational cost of ISF-based decoding}
\label{ssec_dec}
\vspace{-0.50mm}
%
We employ the standard label (token) synchronous beam search
as the decoding algorithm
\cite{Watanabe_IEEEACMTASLP2017,ESPnet}.
As described above,
it is obvious that
the computational cost of ISF
is high,
and we introduce a two-step pruning scheme into the decoding algorithm.
Let $B$ be the beam size and $V$ be the vocabulary size
(number of unique tokens,
$V\hspace{0.11mm}{\gg}\hspace{0.11mm}B$).
At every timestep $t$,
each of
the
$B$ partial hypotheses can be extended with
$V$ tokens,
i.e.,
there are $B$ $\times$ $V$ possible partial hypotheses.
Since
it is difficult to perform ISF
for all of these $B$ $\times$ $V$ hypotheses,
we perform the first pruning
for these hypotheses
based on
the scores obtained from the E2E ASR model
and SF using the FLM
to
obtain the most probable $B$ $\times$ $B$ hypotheses.
%
We then
perform the second pruning
for these $B$ $\times$ $B$ hypotheses
based on the scores obtained from ISF using the BLM
to
obtain the most probable $B$ hypotheses,
which are sent for processing
at the next timestep.
Note that,
with this label synchronous beam search,
the lengths (number of tokens) of 
partial hypotheses are the same during decoding,
and we can introduce batch calculation
to efficiently perform ISF
for the $B$ $\times$ $B$ hypotheses.

Another
idea to reduce the computational cost
is to perform ISF not at every timestep
but at every $I$ ($I \geq 1$) timestep,
as shown in Fig.~\ref{fig_isf}.
In this case,
the BLM score shown in Eq.~(\ref{eq_isf_last})
is replaced by the following score,
\begin{fleqn}[5.00mm]
\begin{eqnarray}
\lefteqn{\log{\Pblm(\left\{\eos,w_{t-I:1},\sos\right\})}}
\nonumber \\
&=&
\log{\Pblm(\eos)}
\nonumber \\
& &
+
\log{{\textstyle\prod_{\tau=t-I}^{1}}
\Pblm(w_{\tau}|\left\{\eos,w_{t-I:\tau+1}\right\})}
\nonumber \\
& &
+
\log{\Pblm(\sos|\left\{\eos,w_{t-I:1}\right\})},
\label{eq_isf_last_iv}
\end{eqnarray}
\end{fleqn}
and the computational cost of ISF is reduced by a factor of $I$.

As described in Section~\ref{sec_intro},
we expect that,
by performing ISF,
we can prevent
incorrect early pruning of
prospective hypotheses.
%
In other words,
to keep the prospective hypotheses alive during decoding,
it may be
more
important
to perform ISF at the earlier stage of decoding
rather than performing it at the later stage of decoding.
To confirm this
prediction
and to reduce the computational cost,
as shown on the top of Fig.~\ref{fig_isf},
we perform ISF only
during
a partial hypothesis is not longer than $L$ tokens
(excluding $\sos$ from the count)
and do not perform ISF
after the hypothesis gets longer than $L$ tokens
until the last post-processing step
(i.e., we perform post-processing
even when we limit the length $L$).
%

In preliminary experiments,
we confirmed that,
even though
applying
the above three cost reduction methods,
the computational cost of ISF
remains
relatively
high.
%
%
%
Further reduction of the computational cost of ISF is our future work.
In this study,
we
investigate
in Section~\ref{sec_exp}
the effect of ISF on reducing word error rates (WERs)
by changing
the
interval $I$
and limiting
the
length
$L$.
%
\begin{figure}[t!]
  \centering
  \includegraphics[width=0.880\linewidth]{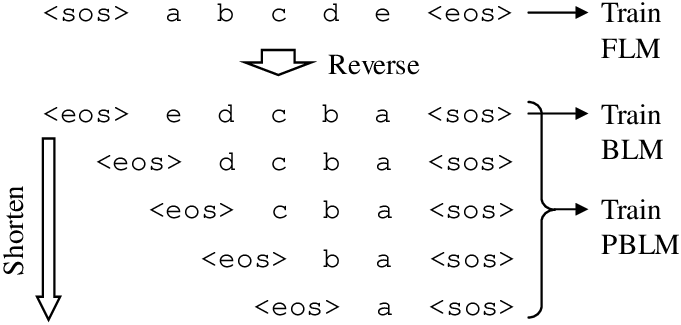}
  \vspace{-1.50mm}
  \caption{\ftblskip Reversed text data including partial sentences (token sequences) considering the framework of ISF used to train a PBLM.}
  \label{fig_pblm}
  \vspace{-2.75mm}
\end{figure}
%
\vspace{-2.50mm}
\subsection{Partial sentence-aware BLM (PBLM) suitable for ISF}
\label{ssec_pblm}
\vspace{-1.00mm}
%
As described in Section~\ref{ssec_fw} and shown in Fig.~\ref{fig_isf},
ISF can be started from a token at any timestep in an ASR hypothesis.
This
indicates
that a normal BLM,
which is trained using
reversed text data,
is not necessarily suitable for ISF,
since
it models token sequences
starting from the end of complete sentences,
whereas we need to apply it to partial hypotheses.

To enhance the effectiveness of ISF,
we propose
using
a PBLM
that can be
more suitable for ISF
than the normal BLM.
As shown in Fig.~\ref{fig_pblm},
we prepare text data for training the PBLM
considering the framework of ISF.
%
We first reverse a sentence (used to train an FLM)
and obtain the reversed sentence (used to train the normal BLM).
Then,
we shorten the reversed sentence
step-by-step
by removing the last token from the sentence at each step
(i.e., interval $I$ is set at 1)
and obtain reversed partial sentences used to train the PBLM.
%
The text data
used to train the PBLM
is greatly augmented compared with those used to train the FLM
or the normal BLM.
We
experimentally
investigate which of the normal BLM or the PBLM is more suitable
for ISF in Section~\ref{sec_exp}.
%
\vspace{-2.60mm}
\section{Relation to prior work}
\label{sec_prior}
\vspace{-1.25mm}
%
As described in Section~\ref{sec_intro},
the proposed ISF is an extension of SF
\cite{Gulcehre_arXiv2015,Bahdanau_ICASSP2016,Chorowski_IS2017,Hori_IS2017,Kannan_ICASSP2018,Toshniwal_SLT2018}
and,
to the best of our knowledge,
this is the first study to use a BLM in SF.
By performing ISF along with SF using an FLM,
we intend to obtain performance improvement
based on
the complementarity of the FLM and
the
BLM,
whose effectiveness has been confirmed in
many studies for
post-processing
($N$-best and lattice rescoring)
of ASR hypotheses
\cite{Xiong_arXiv2017,Xiong_IEEEACMTASLP2017,Irie_IS2018,Kanda_CHiME5,Arora_CHiME6,Ogawa_ICASSP2022}.
In
\cite{Jung_arXiv2021},
the authors
use a bidirectional
LM
for
connectionist temporal classification (CTC)
\cite{Graves_ICML2006}
based
bidirectional
greedy decoding.
In contrast,
in this study,
we use both the FLM and BLM individually
for unidirectional decoding.

This study is inspired by
\cite{Imai_ICASSP2000},
in which the authors developed
a real-time closed captioning system
for
broadcast news
based on a two-pass ASR decoder.
%
The system
progressively outputs
partial ASR results
during decoding
(i.e., streaming).
There are
many
differences between
\cite{Imai_ICASSP2000}
and our study,
such as
their
aims
(streaming ASR
allowing a slight performance degradation
vs. ASR performance improvement)
and
their model
architectures
(traditional GMM/HMM-based vs. recent NN-based).
However,
these two studies
share the concept of
performing
rescoring
during decoding with a fixed interval.
%
\begin{table}[t]
\vspace{-3.00mm}
\caption{\ftblskip Details of the TED-LIUM2 train/dev/test datasets.}
\vspace{1.00mm}
\label{tab_ted2}
\begin{center}
\begin{tabular}{lrrr}
\hline
                           &   Train &   Dev &  Test \\ \hline
Hours                      &   210.6 &   1.6 &   2.6 \\
\# utterances (sentences)  &   92791 &   507 &  1155 \\
\# words                   & 2210368 & 17783 & 27500 \\
\# tokens (subwords)       & 4208823 & 34160 & 51179 \\
Average length (\# tokens) &    45.4 &  67.4 &  44.3 \\ \hline
\end{tabular}
\end{center}
\vspace{-3.25mm}
\caption{\ftblskip Text datasets to train an external FLM, BLM, and PBLM.}
\vspace{-2.75mm}
\label{tab_text}
\begin{center}
\begin{tabular}{lrr}
\hline
                           & FLM/BLM &  PBLM \\ \hline
\# sentences               &     13M &  495M \\
\# tokens (subwords)       &    482M &   14G \\
Average length (\# tokens) &    37.1 &  27.6 \\ \hline
\end{tabular}
\end{center}
\vspace{-6.15mm}
\end{table}
%
\vspace{-3.50mm}
\section{Experiments}
\label{sec_exp}
\vspace{-1.15mm}
%
We conducted experiments based on an ESPnet ASR recipe
using the TED-LIUM2 corpus
\cite{ESPnet,Rousseau_LREC2014}.
%
We used an AED ASR model
\cite{Watanabe_IEEEACMTASLP2017,Guo_ICASSP2021}
consisting
of a Conformer-based encoder
\cite{Gulati_IS2020}
and a Transformer-based decoder
\cite{Vaswani_NIPS2017}
(and also a CTC
\cite{Graves_ICML2006}
module)
as the E2E ASR model
and LSTMLMs
\cite{Hochreiter_NeuralComput1997,Sundermeyer_IS2012}
as the external LMs.
%
Details of
the model structures and the training/decoding settings
can be found in the recipe
\cite{ESPnet}.
We used PyTorch
\cite{Paszke_NeurIPS2019}
for all the NN modeling in this study.
%
\vspace{-2.25mm}
\subsection{Experimental settings}
\label{ssec_set}
\vspace{-0.75mm}
%
Table~\ref{tab_ted2} shows
the details
of the TED-LIUM2
training,
development,
and test datasets
\cite{Rousseau_LREC2014}.
We trained
a SentencePiece model
\cite{Kudo_EMNLP2018}
using the word-based text training data
and tokenized the word-based text training, development, and test datasets
to obtain their token (subword)-based versions.
%
We set the vocabulary size
(number of unique tokens)
$V$ at 500.
Using the training data,
we trained the AED ASR model
(\# encoder / decoder layers $=$ $17$ / $4$,
\# encoder / decoder units $=$ $1024$ / $1024$,
\# attention dimensions / heads $=$ $512$ / $4$)
\cite{Watanabe_IEEEACMTASLP2017,Guo_ICASSP2021}.
%
We did not perform speed perturbation
\cite{Ko_IS2015}
and SpecAugment
\cite{Park_IS2019}
in this study.

Table~\ref{tab_text} shows
the text datasets to train an external FLM, BLM, and PBLM.
We prepared the PBLM training data
with the procedure described in Section~\ref{ssec_pblm},
which resulted in a significant data augmentation.
%
%
All the LMs are the LSTMLMs
\cite{Hochreiter_NeuralComput1997,Sundermeyer_IS2012}
sharing the same structure,
i.e.,
four LSTM layers
(each of them has 2048 units)
and a softmax output layer of the vocabulary size
($V$ $=$ $500$).
We trained these LMs using each of the training datasets
with the
stochastic gradient descent
(SGD)
optimizer for two epochs.

Using the AED ASR model and the three external LMs,
we performed ASR based on the decoding algorithm
described in Section~\ref{ssec_dec}.
We set the beam size $B$ at ten
and optimized the three weighting factors
($\alpha$, $\beta$, $\gamma$) shown in Eq.~(\ref{eq_isf})
using the development data.
We set them
at (0.5, 0.0, 2.0) when we performed SF using the FLM,
at (0.0, 0.5, 2.0) when we performed ISF using the BLM or PBLM,
and at (0.5, 0.5, 5.0) when we performed both SF and ISF.
We performed decoding on a token basis
but evaluated the ASR performance
with WERs.
%
\vspace{-2.25mm}
\subsection{LM evaluation based on perplexities}
\label{ssec_res_ppl}
\vspace{-0.75mm}
%
Before showing the ASR results,
we
show the LM evaluation results
with token-based
perplexities.
To calculate perplexities,
we prepared two types of text datasets.
One is the
standard
dataset
that contains
only complete sentences,
and the other
is
that
also includes
partial sentences,
as described in Section~\ref{ssec_pblm}
and shown in Fig.~\ref{fig_pblm}.

Table~\ref{tab_ppl} shows
the token-based
development and test data perplexities obtained with
the FLM, BLM, and PBLM.
Considering the vocabulary size ($V\!=\!500$),
these perplexity values are reasonable.
We can confirm that
the BLM shows slightly better
performance
than the PBLM
for the ``only complete'' sentences case,
in contrast,
the PBLM shows slightly better
performance
than the BLM
for the ``including partial'' sentences case.
%
These results suggest that
the PBLM
would
be more suitable for ISF than the BLM.
%
\begin{table}[t]
\caption{\ftblskip Token-based dev/test data (only complete / including partial sentences) perplexities obtained with the FLM, BLM, and PBLM.}
\vspace{-3.00mm}
\label{tab_ppl}
\begin{center}
\begin{tabular}{rcccc}
\hline
& \multicolumn{2}{c}{Only complete}
& \multicolumn{2}{c}{Including partial} \\
Model
& \makebox[8.0mm]{Dev}
& \makebox[8.0mm]{Test}
& \makebox[8.0mm]{Dev}
& \makebox[8.0mm]{Test}
\\ \hline
FLM   & 10.2 & 11.8 &  --- &  --- \\
BLM   & 10.1 & 11.8 & 15.1 & 17.1 \\
PBLM  & 10.6 & 12.2 & 14.7 & 16.0 \\ \hline
\end{tabular}
\end{center}
\vspace{-4.50mm}
\end{table}
%
\vspace{-1.25mm}
\subsection{Comparison of SF methods and effect of interval}
\label{ssec_res_iv}
\vspace{-0.00mm}
%
Table~\ref{tab_res_iv} shows 
the
comparison results of the SF methods
and the effect of interval $I$
for applying ISF
(Section~\ref{ssec_dec} and Fig.~\ref{fig_isf}).
%
We set interval $I$ at 1, 2, 5, 10, and $\infty$.
Here,
$I\!=\!\infty$ means that we do not perform ISF for partial hypotheses
but perform it only for completed hypotheses as a kind of post-processing
(Section~\ref{ssec_fw} and Fig.~\ref{fig_isf}).
%
First,
we can confirm that SF using the FLM
(hereafter
referred to as SF-FLM)
steadily reduces the WERs
from the baseline that does not perform SF (W/o SF)
as reported in the previous studies
\cite{Tuske_IS2020,Tuske_IS2021,Gulcehre_arXiv2015,Bahdanau_ICASSP2016,Chorowski_IS2017,Hori_IS2017,Kannan_ICASSP2018,Toshniwal_SLT2018,Sriram_IS2018,Shan_ICASSP2019,McDermott_ASRU2019}.

Next,
by comparing the results of ISF-PBLM
(methods 7 to 11)
with those of ISF-BLM
(methods 2 to 6),
ISF-PBLM performs better on the development data
and comparably or slightly better on the test data.
This improvement is supported by the better modeling capability
of the PBLM for partial hypotheses,
as we confirmed in Section~\ref{ssec_res_ppl}.
We can
also
confirm
that
ISF-PBLM
performs comparably
with SF-FLM.
As regards interval $I$,
smaller values
(e.g., 1 or 2)
are better than larger values
(except for the results of ISF-BLM for the development data).
From these results,
we can confirm that it is important to perform ISF not only
for completed hypotheses as post-processing
but also for partial hypotheses during decoding.
%

Finally,
we can confirm that,
by combining SF-FLM and ISF-PBLM
(methods 12 to 16, SF-FLM+ISF-PBLM),
we can obtain further performance improvement
compared with
when
performing
SF or ISF individually.
This
improvement is achieved
thanks to the complementarity of the FLM and PBLM
\cite{Xiong_arXiv2017,Xiong_IEEEACMTASLP2017,Irie_IS2018,Kanda_CHiME5,Arora_CHiME6,Ogawa_ICASSP2022}.
Also in this case,
we can
confirm
the importance of performing ISF
for partial hypotheses during decoding.
SF-FLM+ISF-PBLM with $I\!=\!2$ achieves about 8\%
relative WER reductions from
SF-FLM,
and these reductions are statistically significant at the 1\% level.
%
\vspace{-1.25mm}
\subsection{Effect of limiting partial hypothesis length to apply ISF}
\label{ssec_res_hl}
\vspace{-0.00mm}
%
As described in Section~\ref{ssec_dec}
and shown in Fig.~\ref{fig_isf},
we investigated the effect of
limiting
the
partial hypothesis length $L$
(i.e.,
the first $L$ tokens)
to apply ISF-PBLM.
%
We employed method 12,
shown in Table~\ref{tab_res_iv},
i.e., SF-FLM+ISF-PBLM with interval $I\!=\!1$.
By setting
$I$ at 1,
the number of times we apply ISF becomes
equal to the length of the partial hypothesis.
We also employed
SF-FLM+ISF-PBLM with
$I\!=\!1$
but without performing post-processing
as another version of method 12,
i.e., method
12${}^{\prime}$,
to exclude
(and also to confirm)
the effect of post-processing.
We conducted experiments on the development data
by setting $L$ at 10, 20, 30, 50, and $\infty$,
since the average length (number of tokens) of
the
utterances
in the development data is 67.4,
as shown in Table~\ref{tab_ted2}.
Here,
$L\!=\!\infty$
means that we do not limit the length,
i.e.,
the default setting.

\begin{figure}[t!]
  \centering
  \includegraphics[width=0.997\linewidth]{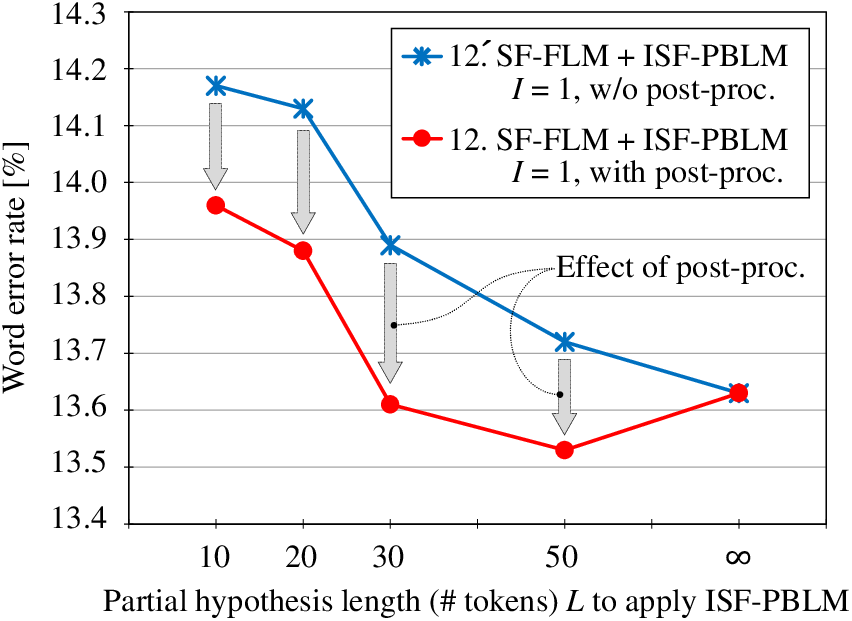}
  \vspace{-5.50mm}
  \caption{\ftblskip Effect of limiting the partial hypothesis length $L$ (\# tokens) to apply ISF-PBLM and that of post-processing on the dev data.}
  \label{fig_hl}
  \vspace{-3.15mm}
\end{figure}

Figure~\ref{fig_hl} shows WERs as a function of
the partial hypothesis length $L$
to apply ISF-PBLM.
%
As can be seen in the result of method 12${}^{\prime}$,
even with
$L\!=\!10$,
we can obtain a steady WER reduction
from the SF-FLM baseline (14.6\%).
At the same time,
we can confirm that,
contrary to our
prediction
described in Section~\ref{ssec_dec},
performing ISF at the later stage of decoding
is also important
as performing it at the earlier stage,
since the WER is gradually reduced by lengthening $L$.
However,
from the result of method 12,
we can confirm that,
with the help of post-processing,
we can
limit $L$
by 30 (tokens),
i.e., about half the length of the complete
sentence
on average.
This result supports our
prediction
to some extent.
%
\begin{table}[t]
\caption{\ftblskip Comparison of the SF methods and the effect of the interval for applying ISF on the development and test datasets.}
\vspace{-3.00mm}
\label{tab_res_iv}
\begin{center}
\begin{tabular}{rlrrr}
\hline
No.  & Method      & Interval ($I$) &      Dev   &      Test  \\ \hline
0.   & W/o SF            &      --- &      15.4  &      11.2  \\ \hline
1.   & SF-FLM            &      --- &      14.6  &       9.9  \\ \hline
2.   & ISF-BLM           &        1 &      15.1  &      10.1  \\
3.   &                   &        2 &      14.9  &      10.2  \\
4.   &                   &        5 &      15.0  &      10.3  \\
5.   &                   &       10 &      15.1  &      10.3  \\
6.   &   & $\infty$\,\,(post-proc.) &      14.8  &      10.5  \\ \hline
7.   & ISF-PBLM          &        1 &      14.2  &      10.1  \\
8.   &                   &        2 &      14.2  &      10.0  \\
9.   &                   &        5 &      14.4  &      10.2  \\
10.  &                   &       10 &      14.7  &      10.3  \\
11.  &   & $\infty$\,\,(post-proc.) &      14.8  &      10.5  \\ \hline
12.  & SF-FLM + ISF-PBLM &        1 &      13.6  &       9.2  \\
13.  &                   &        2 & {\bf 13.5} &  {\bf 9.1} \\
14.  &                   &        5 &      14.0  &       9.2  \\
15.  &                   &       10 &      13.8  &       9.3  \\
16.  &   & $\infty$\,\,(post-proc.) &      13.9  &       9.4  \\ \hline
\end{tabular}
\end{center}
\vspace{-3.00mm}
\end{table}
%
\vspace{-1.35mm}
\section{Conclusion and future work}
\vspace{-1.00mm}
%
We proposed a new shallow fusion method,
i.e.,
iterative shallow fusion (ISF),
using a backward LM for E2E ASR
and showed its effectiveness experimentally.
Future work will include
further
computational cost reduction of ISF
(Section~\ref{ssec_dec}),
evaluation using other ASR tasks,
implementation
on
other E2E ASR models
\cite{Graves_ICML2012,Saon_ICASSP2020,Moriya_ICASSP2022},
and combination with other LM integration methods
\cite{McDermott_ASRU2019,Meng_SLT2021,Moriya_ICASSP2022}.
%
\baselineskip 10.14pt
\bibliographystyle{IEEEtran_ogawa}
\bibliography{ogawa}

\begin{thebibliography}{10}
\providecommand{\url}[1]{#1}
\csname url@samestyle\endcsname
\providecommand{\newblock}{\relax}
\providecommand{\bibinfo}[2]{#2}
\providecommand{\BIBentrySTDinterwordspacing}{\spaceskip=0pt\relax}
\providecommand{\BIBentryALTinterwordstretchfactor}{4}
\providecommand{\BIBentryALTinterwordspacing}{\spaceskip=\fontdimen2\font plus
\BIBentryALTinterwordstretchfactor\fontdimen3\font minus
  \fontdimen4\font\relax}
\providecommand{\BIBforeignlanguage}[2]{{%
\expandafter\ifx\csname l@#1\endcsname\relax
\typeout{** WARNING: IEEEtran.bst: No hyphenation pattern has been}%
\typeout{** loaded for the language `#1'. Using the pattern for}%
\typeout{** the default language instead.}%
\else
\language=\csname l@#1\endcsname
\fi
#2}}
\providecommand{\BIBdecl}{\relax}
\BIBdecl

\bibitem{Xiong_arXiv2017}
W.~Xiong \emph{et~al.}, ``Achieving human parity in conversational speech
  recognition,'' \emph{arXiv:1610.05256v2 [cs.CL]}.

\bibitem{Xiong_IEEEACMTASLP2017}
------, ``Toward human parity in conversational speech recognition,''
  \emph{IEEE/ACM Transactions on Audio, Speech, and Language Processing},
  vol.~25, no.~12, pp. 2410--2423, Dec. 2017.

\bibitem{Saon_IS2017}
G.~Saon \emph{et~al.}, ``English conversational telephone speech recognition by
  humans and machines,'' in \emph{Proc. Interspeech}, 2017, pp. 132--136.

\bibitem{Tuske_IS2020}
Z.~T{\"u}ske, G.~Saon, K.~Audhkhasi, and B.~Kingsbury, ``Single headed
  attention based sequence-to-sequence model for state-of-the-art results on
  {S}witchboard,'' in \emph{Proc. Interspeech}, 2020, pp. 551--555.

\bibitem{Tuske_IS2021}
Z.~T{\"u}ske, G.~Saon, and B.~Kingsbury, ``On the limit of {E}nglish
  conversational speech recognition,'' in \emph{Proc. Interspeech}, 2021, pp.
  2062--2066.

\bibitem{Gulcehre_arXiv2015}
C.~Gulcehre \emph{et~al.}, ``On using monolingual corpora in neural machine
  translation,'' \emph{arXiv:1503.03535v2 [cs.CL]}.

\bibitem{Bahdanau_ICASSP2016}
D.~Bahdanau, J.~Chorowski, D.~Serdyuk, P.~Brakel, and Y.~Bengio, ``End-to-end
  attention-based large vocabulary speech recognition,'' in \emph{Proc.
  ICASSP}, 2016, pp. 4945--4949.

\bibitem{Chorowski_IS2017}
J.~Chorowski and N.~Jaitly, ``Towards better decoding and language model
  integration in sequence to sequence models,'' in \emph{Proc. Interspeech},
  2017, pp. 523--527.

\bibitem{Hori_IS2017}
T.~Hori, S.~Watanabe, Y.~Zhang, and W.~Chan, ``Advances in joint
  {CTC}-attention based end-to-end speech recognition with a deep {CNN} encoder
  and {RNN}-{LM},'' in \emph{Proc. Interspeech}, 2017, pp. 949--953.

\bibitem{Kannan_ICASSP2018}
A.~Kannan, Y.~Wu, P.~Nguyen, T.~N. Sainath, Z.~Chen, and R.~Prabhavalkar, ``An
  analysis of incorporating an external language model into a
  sequence-to-sequence model,'' in \emph{Proc. ICASSP}, 2018, pp. 5824--5828.

\bibitem{Toshniwal_SLT2018}
S.~Toshniwal, A.~Kannan, C.-C. Chiu, Y.~Wu, T.~N. Sainath, and K.~Livescu, ``A
  comparison of techniques for language model integration in encoder-decoder
  speech recognition,'' in \emph{Proc. SLT}, 2018, pp. 369--375.

\bibitem{Sriram_IS2018}
A.~Sriram, H.~Jun, S.~Satheesh, and A.~Coates, ``Cold fusion: {T}raining
  {S}eq2{S}eq models together with language models,'' in \emph{Proc.
  Interspeech}, 2018, pp. 387--391.

\bibitem{Shan_ICASSP2019}
C.~Shan \emph{et~al.}, ``Component fusion: {L}earning replaceable language
  model component for end-to-end speech recognition system,'' in \emph{Proc.
  ICASSP}, 2019, pp. 5631--5635.

\bibitem{McDermott_ASRU2019}
E.~McDermott, H.~Sak, and E.~Variani, ``A density ratio approach to language
  model fusion in end-to-end automatic speech recognition,'' in \emph{Proc.
  ASRU}, 2019, pp. 434--441.

\bibitem{Meng_SLT2021}
Z.~Meng \emph{et~al.}, ``Internal language model estimation for domain-adaptive
  end-to-end speech recognition,'' in \emph{Proc. SLT}, 2021, pp. 243--250.

\bibitem{Moriya_ICASSP2022}
T.~Moriya \emph{et~al.}, ``Hybrid {RNN}-{T}/{A}ttention-based streaming {ASR}
  with triggered chunkwise attention and dual internal language model
  integration,'' in \emph{Proc. ICASSP}, 2022, pp. 8282--8286.

\bibitem{Irie_IS2018}
K.~Irie, Z.~Lei, L.~Deng, R.~Schl{\"{u}}ter, and H.~Ney, ``Investigation on
  estimation of sentence probability by combining forward, backward and
  bi-directional {LSTM}-{RNN}s,'' in \emph{Proc. Interspeech}, 2018, pp.
  392--395.

\bibitem{Kanda_CHiME5}
N.~Kanda \emph{et~al.}, ``The {H}itachi/{JHU} {CH}i{ME}-5 system: {A}dvances in
  speech recognition for everyday home environments using multiple microphone
  arrays,'' in \emph{Proc. of The 5th Intl. Workshop on Speech Processing in
  Everyday Environments (CHiME 2018)}, 2018.

\bibitem{Arora_CHiME6}
A.~Arora \emph{et~al.}, ``The {JHU} multi-microphone multi-speaker {ASR} system
  for the {CH}i{ME}-6 challenge,'' in \emph{Proc. of The 6th Intl. Workshop on
  Speech Processing in Everyday Environments (CHiME 2020)}, 2020.

\bibitem{Ogawa_ICASSP2022}
A.~Ogawa, N.~Tawara, M.~Delcroix, and S.~Araki, ``Lattice rescoring based on
  large ensemble of complementary neural language models,'' in \emph{Proc.
  ICASSP}, 2022, pp. 6517--6521.

\bibitem{Rousseau_LREC2014}
A.~Rousseau, P.~Del{\'e}glise, and Y.~Est{\`e}ve, ``Enhancing the {TED}-{LIUM}
  corpus with selected data for language modeling and more {TED} talks,'' in
  \emph{Proc. LREC}, 2014, pp. 3935--3939.

\bibitem{Watanabe_IEEEACMTASLP2017}
S.~Watanabe, T.~Hori, S.~Kim, J.~R. Hershey, and T.~Hayashi, ``Hybrid
  {CTC}/{Attention} architecture for end-to-end speech recognition,''
  \emph{IEEE/ACM Transactions on Audio, Speech, and Language Processing},
  vol.~11, no.~8, pp. 1240--1253, Dec. 2017.

\bibitem{Guo_ICASSP2021}
P.~Guo \emph{et~al.}, ``Recent developments on {ESP}net toolkit boosted by
  {C}onformer,'' in \emph{Proc. ICASSP}, 2021, pp. 5874--5878.

\bibitem{ESPnet}
S.~Watanabe, ``{ESP}net: {End}-to-end speech processing toolkit,''
  https:{\slash}{\slash}github.com{\slash}espnet{\slash}espnet.

\bibitem{Hochreiter_NeuralComput1997}
S.~Hochreiter and J.~Schmidhuber, ``Long short-term memory,'' \emph{Neural
  Computation}, vol.~9, no.~8, pp. 1735--1780, Nov. 1997.

\bibitem{Sundermeyer_IS2012}
M.~Sundermeyer, R.~Schl{\"u}ter, and H.~Ney, ``{LSTM} neural networks for
  language modeling,'' in \emph{Proc. Interspeech}, 2012.

\bibitem{Jung_arXiv2021}
N.~Jung, G.~Kim, and H.-G. Kim, ``Back from the future: {B}idirectional {CTC}
  decoding using future information in speech recognition,''
  \emph{arXiv:2110.03326v1 [cs.CL]}.

\bibitem{Graves_ICML2006}
A.~Graves, S.~Fern{\'a}ndez, F.~Gomez, and J.~Schmidhuber, ``Connectionist
  temporal classification: {L}abelling unsegmented sequence data with recurrent
  neural networks,'' in \emph{Proc. ICML}, 2006, pp. 369--376.

\bibitem{Imai_ICASSP2000}
T.~Imai, A.~Kobayashi, S.~Sato, H.~Tanaka, and A.~Ando, ``Progressive 2-pass
  decoder for real-time broadcast news captioning,'' in \emph{Proc. ICASSP},
  2000, pp. 1599--1562.

\bibitem{Gulati_IS2020}
A.~Gulati \emph{et~al.}, ``Conformer: {C}onvolution-augmented {T}ransformer for
  speech recognition,'' in \emph{Proc. Interspeech}, 2020, pp. 5036--5040.

\bibitem{Vaswani_NIPS2017}
A.~Vaswani \emph{et~al.}, ``Attention is all you need,'' in \emph{Proc. NIPS},
  2017, pp. 5998--6008.

\bibitem{Paszke_NeurIPS2019}
A.~Paszke \emph{et~al.}, ``{P}y{T}orch: {A}n imperative style, high-performance
  deep learning library,'' in \emph{Proc. NeurIPS}, 2019, pp. 8024--8035.

\bibitem{Kudo_EMNLP2018}
T.~Kudo and J.~Richardson, ``{S}entence{P}iece: {A} simple and language
  independent subword tokenizer and detokenizer for neural text processing,''
  in \emph{Proc. EMNLP}, 2018, pp. 66--71.

\bibitem{Ko_IS2015}
T.~Ko, V.~Peddinti, D.~Povey, and S.~Khudanpur, ``Audio augmentation for speech
  recognition,'' in \emph{Proc. Interspeech}, 2015, pp. 3586--3589.

\bibitem{Park_IS2019}
D.~S. Park \emph{et~al.}, ``{S}pec{A}ugment: {A} simple data augmentation
  method for automatic speech recognition,'' in \emph{Proc. Interspeech}, 2019,
  pp. 2613--2617.

\bibitem{Graves_ICML2012}
A.~Graves, ``Sequence transduction with recurrent neural networks,'' in
  \emph{Proc. ICML}, 2012.

\bibitem{Saon_ICASSP2020}
G.~Saon, Z.~T{\"u}ske, and K.~Audhkhasi, ``Alignment-length synchronous
  decoding for {RNN} transducer,'' in \emph{Proc. ICASSP}, 2020, pp.
  7804--7808.

\end{thebibliography}

\end{document}